**Article type:** (**Original Paper**)

**Title** (Control of spectral extreme events in ultrafast fibre lasers by a genetic algorithm)

*Xiuqi Wu[1], Ying Zhang[1], Junsong Peng[1,2,\*], Sonia Boscolo[3], Christophe Finot[4], and Heping Zeng[1,5,\*\*]*

\*Corresponding Author: jspeng@phy.ecnu.edu.cn; hpzeng@phy.ecnu.edu.cn

[1]State Key Laboratory of Precision Spectroscopy, East China Normal University, Shanghai 200062, China
[2]Collaborative Innovation Center of Extreme Optics, Shanxi University, Taiyuan, Shanxi 030006, China
[3]Aston Institute of Photonic Technologies, Aston University, Aston Triangle, Birmingham B4 7ET, UK
[4]Laboratoire Interdisciplinaire Carnot de Bourgogne, UMR 6303 CNRS – Université de Bourgogne Franche-Comté, F-21078 Dijon Cedex, France
[5]Chongqing Key Laboratory of Precision Optics, Chongqing Institute of East China Normal University, Chongqing 401120, China

Abstract: Extreme wave events or rogue waves (RWs) are both statistically rare and of exceptionally large amplitude. They are observed in many complex systems ranging from oceanic and optical environments to financial models and Bose-Einstein condensates. As they appear from nowhere and disappear without a trace, their emergence is unpredictable and non-repetitive, which make them particularly challenging to control. Here, we extend the use of genetic algorithms (GAs), which have been exclusively designed for searching and optimising stationary or repetitive processes in nonlinear optical systems, to the active control of extreme events in a fibre laser cavity.

Feeding real-time spectral measurements into a GA controlling the electronics to optimise the cavity parameters, we are able to trigger wave events in the cavity that have the typical statistics of RWs in the frequency domain. This accurate control enables the generation of the optical RWs with a spectral peak intensity 32.8 times higher than the significant intensity threshold. A rationale is proposed and confirmed by numerical simulations of the laser model for the related frequency up- and down-shifting of the optical spectrum that are experimentally observed.

## 1. Introduction

First introduced in the context of oceanic waves,[1] the concept of extreme events or rogue waves (RWs), i.e., statistically-rare giant-amplitude waves, has been transferred to other natural environments such as the atmosphere,[1] as well as to the solid grounds of research laboratories. Observations of extreme events are now ubiquitous in various fields of science including plasma physics,[2, 3] condensed matter physics,[4] superfluidity and nonlinear optics.[5-7] In particular, fibre-optics systems provide ideal test-beds for investigating the evolution dynamics of RWs[7] and the conditions for their emergence[8] owing to the possibility of recording a large amount of events in a relatively short time and the relative simplicity and precision of fibre-optics experiments. Since the first observation of optical RWs on a platform of supercontinuum generation in a microstructured optical fibre[5] and thanks to the advances of ultrafast measurements, it has become possible to accurately measure the temporal and spectral characteristics of the nonlinear evolution of such waves and to control the amplitude and phase of the initial conditions fed to the system. Therefore, the occurrence of extreme events in cavity-free optical systems is now well understood and has been intimately associated with the modulation instability process and solitons over a finite background. As noise-driven dynamics are responsible for the appearance of RWs upon supercontinuum evolution, the ensuing dynamics are inherently unpredictable and uncontrollable.

Yet, a well-controlled seed signal can help preserve the coherence of the continuum[6, 9] or can be used to induce collision of breather structures leading to the excitation of enhanced RWs.[10] However, such control methods cannot be implemented in more complex systems, such as those involving optical feedback, where the physical mechanisms underpinning the occurrence of RWs may differ from modulation-instability related dynamics. In this context, ultrafast fibre lasers, which represent interesting realisations of dissipative nonlinear systems with dynamics driven by a complex interplay among the effects of nonlinearity, dispersion and energy exchange,[11] provide excellent platforms to observe optical RWs. These can arise from soliton interactions,[12-15] Raman scattering,[16, 17] noise-like pulse emission,[18, 19] or soliton/breather explosions.[20-22]

These markedly different scenarios call for a universal method to achieve on-demand generation of RWs. Machine-learning strategies, referring to the use of statistical techniques and numerical algorithms to carry out tasks without explicit programmed and procedural instructions and being widely deployed in many areas of science and engineering,[23] appear to provide the appropriate solution. In the field of ultrafast photonics, machine-learning approaches and the use of genetic algorithms (GAs) have recently led to several dramatic improvements in dealing with the multivariable optimisation problem associated with achieving desired operating regimes of fibre lasers.[24-26] Yet, these algorithms have been designed to target regimes of parameter-invariant, stationary pulse generation,[24, 27-36] while the latest advances relate to the generation of non-stationary repetitive patterns, such as breathing solitons and multi-breather complexes,[36] stationary soliton pairs,[37] or noise-like pulses.[38] By contrast, the on-demand generation of extreme waves, which are non-repetitive and rare events, represents a much harder task, and whether machine learning can be used to control such nonequilibrium dynamics widely present in many physical systems remains an open question. It is worthy to note that machine learning has been applied to compute the temporal intensity of optical RWs from spectral measurements only,[39] thereby relieving the experimental burden of direct time-domain observations. However, active control of extreme wave events by machine learning has not been demonstrated to date.

In this paper, we report on the intelligent generation of optical RWs and control of their intensity in a mode-locked fibre laser using GA-based optimisation of the four-parameter intracavity nonlinear transfer function steered by electronically driven polarisation control. We define merit functions relying on the statistical spectral properties of the laser output, which are capable to drive the formation of pulses in the laser displaying substructures with record-high intensities in the frequency domain. In particular, a super RW of intensity 32.8 times higher than the significant intensity threshold is synthesised. These extreme spectral events correlate with extreme variations of the pulse energy. Quite remarkably, significant frequency up- or down-shifting of the optical spectrum is also associated with the emergence of these waves, which is confirmed by numerical simulations of the laser model based on the modified nonlinear Schrödinger equation.

## 2. RESULTS

### 2.1. Experimental setup and measurements

The experimental setup for generating optical RWs, sketched in Fig. 1, is a typical fibre ring cavity in which a 1.3-m-long erbium-doped fibre (EDF) constitutes the gain medium, pumped by a laser diode operating at 980 nm through a wavelength-division multiplexer. Other fibres in the cavity are a section of dispersion-compensating fibre (DCF) and pieces of standard single-mode fibre (SMF) from the pigtails of the optical components used. The group-velocity dispersion (GVD) values ($\beta_2$) of the three fibre types are 65, 62.5, and –22.8 ps$^2$/km, respectively, yielding a net normal dispersion of 0.028 ps$^2$ at the operating wavelength of ~1.5 $\mu$m. The fundamental repetition frequency of the cavity is 16.765 MHz. The mode-locked laser operation is obtained thanks to an effective saturable absorber based on the nonlinear polarisation evolution (NPE) effect.[40] The nonlinear transfer function of the NPE-based mode locking is controlled by an electronically driven polarisation controller (EPC) working together with a polarisation-dependent isolator. The EPC consists of four fibre squeezers oriented at 45° to each other and can generate all possible states of polarisation over the Poincaré sphere, where each set of voltages applied on the squeezers corresponds to a specific state.[27, 31, 33, 41] The overall envelope fluctuations of the laser output are directly measured via a fast photodiode (20-ps response time, 50-GHz bandwidth) connected to a real-time oscilloscope (33-GHz bandwidth, 80-GSa/s sampling rate). Since the photodiode measurements are unable to resolve the details of the temporal structure on the circulating pulses, the real-time shot-to-shot dynamics of the laser output are characterised with the time-stretch dispersive Fourier transform (DFT) method,[42-44] which has been widely used for RW detection,[18, 19] the study of noise-like lasers,[17] as well as the study of transient or unstable laser states.[45-50] In the latter case, the output of the laser generally contains both localised and non-localised waves. While the DFT data carefully reflects the spectral information for the localised waves, this is no longer the case for the non-localised waves. Therefore, simultaneous temporal intensity measurements are generally performed to assist the DFT data analysis and distinguish between localised and non-localised waves. Non-localised light typically arises from relaxation oscillations or quasi-continuous waves.[47, 48, 50-52] The DFT setup consists of a long segment of normally dispersive fibre that provides an accumulated dispersion of $DL \approx -1200$ ps/nm, thereby enabling the mapping of the spectrum of a short laser pulse in the time domain. From the photodetection of the DFT output signal on a fast photodiode, the optical spectrum for each pulse is obtained directly on the oscilloscope, with a resolution of $\delta\lambda = 1/(DL \cdot BW) \approx 0.025$ nm, where $BW$ is the bandwidth of the photodetection. The oscilloscope is connected to a computer that runs the GA and controls the EPC via a field programmable gate array (FPGA) and four digital-to-analogue converters (DACs). During the searching process, the signals generated by the algorithm are delivered to the FPGA, which adjusts the control voltages of the EPC through the DACs. The DACs translate the instructions from the FPGA to the voltages and finally act on the EPC.

**2.2 Genetic algorithm optimisation**

The intelligent search of RWs is realised via an GA whose principle, illustrated in Figs. 1(b) and 1(c), mimics mechanisms inspired by the biological evolution: individuals composing a population progress through successive generations only if they are among the fittest.[53] In our case, an individual is a laser state, associated with the nonlinear transfer function defined by the four control voltages applied on the EPC; these voltages are therefore the genes of the individuals. The process begins with a randomly created population of individuals making up the initial generation. The output of the system is measured for each individual in the generation, evaluated by a user-defined merit function (also known as objective function or fitness) and assigned a score. The GA then creates the next generation by breeding individuals from the preceding generation, with the probability that an individual is selected to be a 'parent' based on their score ('roulette wheel' selection). Two new individuals – children - are created from the crossover of two randomly selected parents, namely the interchange of their genes. A mutation probability is also specified, which can randomly alter the children's genes, thus allowing for the genetic sequence to be refreshed. The process repeats until the algorithm converges and an optimal individual is produced. In our experiment, the initial population is set to 100 individuals, and the next generations have a smaller population composed of 30 individuals (10 parents and 20 children). Evaluation of the properties of an entire generation of individuals typically takes 2.5 minutes.

The merit function is the key ingredient of any GA, as it defines the optimisation target. In our case, in order to connect the merit function to the emergence of spectral RWs, a long time trace of the laser output is recorded by the oscilloscope, which enables through DFT measurements the accumulation of $12\times10^3$ successive single-shot spectra. From these measurements, the distribution of the intensity maxima $I_{MAX}$ for the spectra recorded at each round trip can be computed. We then design merit functions promoting the generation of ordinary and super spectral RWs based on two common statistical criteria for a wave event to be classified as a RW. One is that its height $I_{MAX}$ exceeds twice the significant wave height (SWH) $I_{SWH}$, defined as the mean intensity of the upper third of intensity peaks. Therefore, the merit function for generating RWs is designed as $F_{RW} = C_1 - |C_{RW} - I_{MAX}/I_{SWH}|$, where $C_1$ is a constant set to 35, where this value is determined from the strongest RWs observed in the laser, and $C_{RW}$ is the target intensity relative to the significant intensity $I_{SWH}$. It is straightforward to see that parameter $C_{RW}$ controls both the generaton of RWs and their intensity: if it is set to 2, $F_{RW}$ returns a maximum value ($C_1$) when a RW is generated ($I_{MAX}/I_{SWH} = 2$). Elseways, if it set to a larger value, a RW with proportionately stronger intensity is generated as the GA tunes the laser parameters to make $I_{MAX}/I_{SWH}$ always equals to $C_{RW}$ so as to maximise the fitness.

The other criterium qualifying a wave as a RW is that $I_{MAX}$ be greater than the mean intensity $<I>$ of the intensity peaks by 8 times the standard deviation $\sigma$ from the mean.[54] Super RWs refer to ultra-large deviations from the mean intensity[54, 55] such that $(I_{MAX}-<I>)/\sigma > 32$. Therefore, the merit function for generating super RWs is designed as $F_{SRW} = \alpha(C_2 - |C_{SRW} - I_{MAX}/(<I>+32\sigma)|)$, $\alpha$ and $C_2$ are constants set to 15 and 1.2, respectively ($\alpha$ is used to amplify the fitness), and $C_{SRW}$ is the parameter that controls the formation of super RWs. Based on the super RW defining criterium given above, $C_{SRW}$ is set to a value above 1 to guarantee that $F_{SRW}$ is maximum when a super RW is generated. To control the intensity of super RWs, we have merged the two merit functions $F_{RW}$ and $F_{SRW}$ into a new one: $F_{TSRW} = \alpha(C_2 - |C_{RW} - I_{MAX}/(\bar{I} + 32\sigma)|) + \beta F_{RW}$, where the weights of the two components are determined empirically ($\alpha = 15$, $\beta = 1$).

Examples of GA optimisation curves for ordinary and super RW generation are presented in panels (a) and (b) of Fig. 2, which show the best and average values of the respective ratios $I_{MAX}/I_{SWH}$ and $(I_{MAX} - <I>)/\sigma$, observed within the population for successive generations. We see that in the case of ordinary RWs the best and average $I_{MAX}/I_{SWH}$ values increase and converge to the optimisation target of $C_{RW} = 8\pm10\%$ and an optimised value, respectively, after five generations. For the optimisation time of the GA to be reasonably short (the average time is 25 minutes), the computer program stops once the output strength is within 10% of the target value. The best and average values of the fitness ratio used for generating super RWs display a similar evolution although their convergence is slower. We note that as RWs can be non-stationary processes, an important issue is whether they can be reproduced generation after generation. As we can see from Fig. 2, for both types of RWs some generations after the optimal state has been first reached (blue diamonds between red ones) have smaller values than the target values. Nonetheless, the GA then modifies the genes of the individuals and the laser emits the target RWs again. These results evidence the ability of our GA to "lock" optical RWs: whilst some generations may deviate from the target RWs, the GA can successfully restore the laser to emitting RWs of the required intensity. Such locking ability of the GA has been confirmed by numerous experiments, where additional examples are given in the Supplementary Information (Figs. S1 and S2). Representative evolutions of the intensity maxima $I_{MAX}$ of the photo-detected signals after time stretching (DFT data) over 12,000 successive cavity round trips when the laser operates in the optimal ordinary and super RW modes are depicted in panels (c) and (d) of Fig. 2, respectively. We see that while the emergence of ordinary RWs in the laser cavity is accompanied by a pronounced background which is induced by relaxation oscillations, super RWs feature low fluctuations in the background. Such a difference has been evidenced by several experiments, two examples of which are shown in Figs. S1 and S2 in the Supplementary Information. In this regard, optical super RWs resemble more their oceanic counterparts, which appear unexpectedly and disappear without a trace.[55]

## 3. DISCUSSION

### 3.1 Characteristics of ordinary and super rogue waves

Examples of statistical analysis for the observed ordinary and super RW operations of the laser are provided in Fig. 3. Panels (a) and (b) show the associated distributions of spectral intensity maxima computed from the DFT recordings over 12,000 round trips. Both histograms display a L-shaped profile, with extreme events occurring rarely, yet much more frequently than expected based on the relatively narrow distribution of typical events. However, the occurrence of extreme events with intensities approximately 40 times the average results in a significantly extended tail of super RWs for the distribution of Fig. 3(b), which also displays a smaller standard deviation. Note that contrary to recent works using time-lens measurements,[17, 56] in our experiment we do not have direct access to the real-time time-domain dynamics of the laser, hence we cannot measure the distribution of temporal intensity maxima for successive cavity round trips. Nonetheless, we have recorded the round-trip evolution of the energy of the output pulses calculated by integrating the DFT spectra and plotted it side by side with the corresponding evolution of the spectral intensity maxima (Figs. 3(c) and 3(d)). We can clearly see that for both the ordinary and super RW regimes, the peak intensity fluctuations in frequency domain are tightly connected with changes of the pulse energy, thereby entailing a L-shaped profile of the associated distribution of energy (Figs. 3(e) and 3(f)). The strong link between spectral intensity and energy variations has been confirmed by numerous experiments. Two further examples are provided in the Supplementary Information (Fig. S3 and S4). Therefore, these statistics qualify the extreme events observed as temporal RWs as well. As mentioned in Section 2.1, the DFT signal cannot reflect the spectral information if the temporal trace does not have a localised structure. However, this does not influence the findings of our work. See the discussion in the Supplementary Information (Fig. S3 and S4).

Without active control, the intensities of the RWs generated in many optical systems are generally just above the threshold of 2 times the SWH. For example, RWs of intensities approximately 10 times the SWH have been reported in optical multiple filamentation;[57] Raman scattering in fibre lasers generates RWs with intensities approximately 7 times the SWH.[16] Yet, the intensity of RWs has never been actively controlled in any physical systems to date. Here, the merit functions used by the GA permit to achieve RWs with tailored intensity and to find operating conditions of the laser which enable record-high intensity values. The results obtained with the merit function $F_{RW}$ (see Section 2.2) based on the optimisation of the ratio $I_{MAX}/I_{SWH}$ are summarised in Fig. 4(a). The highest RW so produced has an intensity exceeding the SWH by a factor of 32.8. We also note that whilst the RWs generated by use of the merit function $F_{RW}$ can be very strong, only some of them fulfil the defining criterium of super RWs from which the merit function $F_{SRW}$ is derived. These super RWs are indicated by red circles in Fig. 4(a). In other words, a high $I_{MAX}/I_{SWH}$ ratio is not always synonym of super RWs. Yet, by using the composite merit function $F_{TSRW}$, the GA not only is able to tune the intensity of the RWs, but also to guarantee that they are super RWs, as shown in Fig. 4(b). Quite importantly, Figs. 4(a) and 4(b) together highlight that ordinary and super RWs cannot be distinguished by their intensity only, since the former can have large intensities, e.g., 25 times the SWH (Fig. 4(a)) while the latter can feature relatively small intensities, e.g., only 10 times the SWH (Fig. 4(b)). A second degree of freedom is therefore required to discriminate between ordinary and super RWs, and the standard deviation of the distribution of intensity maxima is fit for purpose. Figure 4(c) shows the distribution of the ordinary and super RWs emitted by the laser in the plane of peak intensity $I_{MAX}$ and standard deviation $\sigma$. A dashed black line separating the regions of existence of ordinary and super RWs is drawn to ease visualisation. We can clearly see that relatively low-intensity events still correspond to super RWs when the standard deviation is quite small. By contrast, high-intensity events do not qualify as super RWs if the standard deviation is too large.

**3.2 Dynamics of the rogue events**

In our laser setup, the automated and high-resolution control of the NPE transfer function enables access to the nonlinear cavity dynamics, and the precise measurements enabled by the DFT technique allow us to track the spectral changes experienced by the circulating pulses in the cavity. This provides a precious tool for controlling the emergence of RWs in the cavity. We are then able to reveal four different generation mechanisms of RWs: Q-switching instability with frequency down- or up-shifting, multiple pulsing, transient noise-like pulsing, and temporal spiking preceding mode locking. We focus here on the frequency shifting process occurring under Q-switching instability operation of the laser, while the other three processes are discussed in the Supplementary Information (Figs. S5, S6 and S10).

Figures 5(a) and 5(d) show typical DFT-recorded evolutions of the intensity maxima $I_{MAX}$ of the laser spectrum over successive cavity roundtrips, and magnified versions of the areas supporting the largest $I_{MAX}$ values are shown in the insets to highlight the development and disappearance of RWs. Since this usual one-dimensional representation of the laser intensity evolution precludes a detailed investigation of the underlying dynamics, a two-dimensional – spatio-spectral – representation [49, 51, 58-60] is used in Figs. 5(b) and 5(e) to capture the transient dynamics of RWs. We can see that the laser spectrum has different structures before and after the occurrence of the RW peak. Indeed, as revealed by the single-shot spectra at the roundtrips that proceed the emergence of the RW peak (Figs. 5(c) and 5(f)), the spectrum features well-defined oscillations, which are quite typical of the spectral broadening of a coherent wave induced by self-phase modulation (SPM) in the fibre. The spectrum extent is up to 10 nm. By contrast, at the round trips succeeding the RW peak emergence, the spectrum structure becomes much more chaotic, seemingly noise-like, with significant changes from round trip to round trip and spanning over 40 nm. Another noticeable feature revealed by the DFT measurements is the progressive and continuous frequency down- or up-shifting of the spectrum that occurs during the rising stage of RWs. Frequency down-shifting due to Raman scattering has been identified as the main wave mechanism of RW formation in supercontinuum generation[5, 6] and one of the possible RW formation mechanisms in mode-locked lasers.[16] However, no significant Raman effect occurs in our laser due to the short length of the cavity (12.26 m). We can see in Fig. 5(c) that in the beginning the frequency components on the long wavelength edge of the spectrum are only slightly stronger than the frequencies on the short-wavelength edge, but then the long-wavelength end of the spectrum grows with increasing pulse energy over the following roundtrips to finally evolve into a RW. As revealed by Fig. 5(f), frequency up-shifting may also be a signature of the development of RWs in the cavity, which confirms our assumption that Raman scattering is not the key physical process driving the observed intra-pulse energy transfer. Both types of frequency shift phenomena shown in Fig. 5 feature a 'winner takes all' scenario: the stronger spectrum end at the start grows even stronger and develops into a spectral RW. To confirm that the DFT measurements are valid in these cases, the corresponding temporal intensity measurements are also provided (Fig. S7 of the Supplementary Information).

To illustrate further these intriguing dynamics, we have performed numerical simulations of the laser using a scalar-field, lumped model that includes the dominant physical effects of the system on the evolution of a pulse over one round trip inside the cavity, namely, GVD and SPM for all the fibres, gain saturation and bandwidth-limited gain for the active fibre,[61] and the discrete effects of a saturable absorber element. The pulse propagation in the optical fibres is modelled by a generalised nonlinear Schrödinger equation, which takes the following form:[62]

$$\psi_z = -\frac{i\beta_2}{2}\psi_{tt} + i\gamma|\psi|^2\psi + \frac{g}{2}\left(\psi + \frac{1}{\Omega^2}\psi_{tt}\right), \quad (1)$$

where $\psi = \psi(z,t)$ is the slowly varying electric field moving at the group velocity along the propoagation coordinate $z$, and $\gamma$ is the Kerr nonlinearity coefficient. We used the nominal experimental GVD values and the estimated nonlinear coefficients of $\gamma = 5\ (\text{W}\cdot\text{km})^{-1}$, $\gamma = 1.1\ (\text{W}\cdot\text{km})^{-1}$ and $\gamma = 5\ (\text{W}\cdot\text{km})^{-1}$ for the EDF, SMF and DCF, respectively. The dissipative terms in Eq. (1) represent linear gain as well a parabolic approximation to the gain profile with the bandwidth $\Omega$. The gain is saturated according to $g(z) = g_0\exp(-E/E_s)$, where $g_0$ is the small-signal gain, which is non-zero only for the gain fibre, $E(z) = \int dt|\psi|^2$ is the pulse energy, and $E_s$ is the gain saturation energy determined by the pump power. The effective nonlinear saturation involved in the NPE mode-locking technique is modelled by an instantaneous and monotonous nonlinear transfer function for the field amplitude: $T = \sqrt{1 - q_0/[1 + P(t)/P_s]}$, where $P(z,t) = |\psi(z,t)|^2$ the instantaneous pulse power. We took as typical values $q_0 = 0.8$ for the unsaturated loss due to the saturable absorber and $P_s = 0.3$ W for the saturation power. Linear losses are imposed after the passive fibre segments, which summarise intrinsic losses and output coupling. The numerical model is solved with a standard symmetric split-step propagation algorithm, and inspired by the experimentally observed asymmetry of the laser spectrum, we have used a picosecond pulsed initial condition with an asymmetric temporal waveform for our model.

The evolution of the spectral intensity over successive cavity round trips illustrated in Figs. 6(a) and 6(b) indicates that the initial asymmetry in time domain is converted into a progressively developing asymmetry in the spectrum, with an intensity peak growing on one edge of the spectrum. These results qualitatively agree with the experimental observations. A red or blue shift of the spectrum occurs depending on the side of the initial temporal asymmetry, thereby showing that the initial condition dictates which spectral components will grow into a RW. The dynamics of RWs induced by frequency up-shifting are presented in the Supplementary Information (Fig. S8). Details of the intracavity spectral dynamics over two roundtrips are provided in Figs. 6(e) and 6(f). We can see that most of the SPM-induced spectral broadening occurs in the EDF, and the temporal asymmetry of the initial pulse results in an energy difference between the short and long wavelength components of the spectrum.[61]

For completeness, the corresponding round-trip evolution of the laser temporal intensity is also plotted in Figs. 6(c) and 6(d). Whereas in the first stages of the evolution the temporal profile of the pulse remains rather unchanged, pronounced oscillations eventually develop on the pulse edge with the larger intensity gradient, which are reminiscent of an intra-pulse shock wave. These oscillations would ultimately lead to pulse collapse and the emergence of noise-like broadband structures similar to those observed in Fig. 5. Therefore, the dynamics of our laser operating at normal net dispersion appear to be rather different from those arising in laser cavities with net anomalous dispersion where the pulse propagation can be destabilised by modulation-instability processes or soliton fission. We should note that while our laser model is able to capture the evolution dynamics relating to the formation of RWs, it cannot reproduce the dynamics connected with their disappearance. A full description of the transient nature of RWs is a particularly challenging task, mainly due to the fact that a Q-switching-type phenomenon is involved in the dynamics as illustrated by Fig. S9 in the Supplementary Information.

Identifying the key parameters of the laser cavity that govern the dynamics of RWs is an important question. Our work shows that these are the parameters of the saturable absorber (NPE) and the pump power. Indeed, when the pump power is around the threshold for stationary mode locking and the parameters of the saturable absorber do not fulfil the stability condition,[63] Q-switching instability disrupts mode locking and leads to the generation of RWs. By contrast, RWs generated by multiple pulsing are observed when the pump power is increased such that it is far beyond the mode locking threshold, as illustrated by Fig. S10 in the Supplementary Information. Intensity distributions of the RWs generated by Q-switching and multiple pulsing are provided in Fig. S11 in the Supplementary Information which shows that Q-switching can promote the generation of RWs with a wide range of intensities (from 5 to 32 $I_{SWH}$) while multiple pulsing induces RWs with smaller intensities within a narrower range (around 7 $I_{SWH}$).

## 4. CONCLUSION

We have demonstrated, for the first time, the possibility of using GAs to promote the emergence and control the intensity of RWs in nonlinear optical systems. Both ordinary and super RWs with tailored intensity can be generated on demand using merit functions that are based on their statistical defining characteristics. This control strategy is general and independent of the physical model considered. In the present work dealing with a fibre laser cavity, we have experimentally demonstrated the handling of extreme spectral events which also correlate with extreme fluctuations of the pulse energy. The real-time capture of a large number of spectral data through DFT measurements along with the guidance provided by numerical simulations of the laser model have suggested a new physical scenario for the emergence of these extreme waves: an initially coherent but asymmetric pulse circulating in the cavity causes one edge of the spectrum to grow and eventually evolve into a RW through the effect of SPM, while a concomitant intra-pulse shock wave develops in the time domain. This shock wave ultimately leads to pulse collapse and the emergence of noise-like broadband structures. By tuning the cavity parameters, the GA can trigger the formation of the initially asymmetric waveform sowing the RW dynamics.

To verify the ability of our control method to promote the emergence of RWs with user-defined intensity in different laser cavities, we have applied the method to a laser with a quite different net cavity dispersion ($-0.09$ ps$^2$). The results are summarised in Fig. S12 of the Supplementary Information, showing that the intelligent system can still control RWs.

The occurrence of RWs is a very general phenomenon in fibre lasers. RWs have been observed in various fibre laser systems, including ring, linear[56, 64] and figure-eight cavities,[16] normal-, anomalous-,[18] and near-zero dispersion[65] cavities, long[15, 66] and short cavities. As mentioned in Section 1, various physical processes have been identified as the main drivers of RW formation in mode-locked fibre lasers. Generally, these processes can be readily accessed through increasing of the laser pump power. Besides, RWs may also appear when turning on a laser. We argue that RWs do not represent a severe threat for fibre lasers mode locked through the NPE technique. Possibly, the inclusion of a physical saturable absorber, such as a semiconductor saturable absorber mirror, in the fibre cavity makes the laser more susceptible to damage. RWs could also be a problem for the detection devices, such as photodiodes, used to monitor the laser emission or for the sample under investigation. Moreover, after the occurrence of a RW event, the laser may switch to very different modes of operation (continuous-wave or non-stationary modes), which is not usually welcomed in the context of applications. To control the nonlinear transfer function of the NPE-based laser mode locking, electronic polarisation controllers have been largely used,[27] [29, 33] featuring a faster control speed than that of liquid-crystal phase retarders.[67, 68] However, RWs have been observed also in fibre laser layouts where polarisation control is not used.[16] It is reasonable to assume that, compatibly with the types of optical elements incorporated in the laser cavity and their offered number of degrees of freedom, it is possible to achieve intelligent control over the generation and intensity of RWs also in fibre lasers not deploying polarisation controllers.

RWs are ubiquitous in nature, and understanding their generation mechanisms in different physical contexts is an important problem.[8, 69] Mode-locked lasers represent ideal testbed systems for studying optical RWs. In this respect, we believe that the conceptually different generation mechanisms of RWs that our machine learning-based control method enables revealing will be of great interest to the laser photonics community.

From the perspective of practical applications, it is crucial that a mode-locked laser remains in the stable mode locking regime without switching to nonstationary operation states such as RWs. Therefore, knowing the conditions for the emergence of RWs is important to avoid such states. Our work indicates that RWs can be stimulated by Q-switching or multiple pulsing in mode-locked fibre lasers, thus one can suppress these two effects to avoid RWs. Q-switching instability can be suppressed by operating the laser at pump powers well above the mode-locking threshold or, alternatively, by deploying a saturable absorber with low modulation depth and small saturation energy.[63] To suppress multiple pulsing, a short fibre cavity can be used.

It is reasonable to expect the machine-learning method used in this work to be applicable to the control of RWs in other nonlinear systems. In particular, it would be of great interest to be able to control RWs in water wave tanks, where the dynamics are closely related to those in the ocean.[70] Besides, RWs refer to unstable states of complex systems. As demonstrated here in the case of a laser system, the use of control algorithms can make these instabilities accessible in a "repetitive" manner so that laborious manual tuning of the system's parameters is no longer required, and this facilitates the exploration of the rich underlying physics. Therefore, our work may open the way to the control and study of instabilities in a wide range of complex systems.

**Supporting Information**

Figure S1: Ordinary RWs just below the intensity threshold for super RWs, displaying weak relaxation oscillations.

Figure S2: Ordinary RWs far from the intensity threshold for super RWs, displaying quite strong relaxation oscillations.

Figure S3: Characteristics of RWs with relaxation oscillations involved in the dynamics.

Figure S4: Characteristics of RWs with no relaxation oscillations involved in the dynamics.

Figure S5: Experimentally observed RW dynamics induced by transient noise-like pulsing.

Figure S6: Experimentally observed RW dynamics induced by temporal spiking preceding mode locking.

Fig. S7. RW dynamics induced by frequency down-shifting and up-shifting in the time and frequency domains.

Figure S8: Dynamics of frequency up-shifting induced RWs as obtained from numerical simulations of the laser model.

Figure S9: Q-switching process involved in the RW dynamics induced by frequency shifting.

Figure S10: Experimentally observed RW dynamics at high pump power (150 mW).

Figure S11: Intensity distributions of the RWs generated by Q-switching and multiple pulsing.

Figure S12: Intelligent control of RWs in a laser cavity with a different net dispersion (–0.09 ps$^2$).

**Acknowledgements**   (National Key Research and Development Program (2018YFB0407100); the National Natural Science Fund of China (11621404, 11561121003, 11727812, 61775059, and 11704123); the Key Project of Shanghai Education Commission (2017-01-07-00-05-E00021); the Science and Technology Innovation Program of Basic Science Foundation of Shanghai (18JC1412000); the Shanghai Rising-Star Program; Sustainedly Supported Foundation by National Key Laboratory of Science and Technology on Space Microwave under Grant 2022-WDKY-SYS-DN-04; UK Engineering and Physical Sciences Research Council (EP/S003436/1); French National Research Agency (ANR-20-CE30-0004).

**Competing interests:** There are no financial competing interests.

**Keywords**: (rogue waves, mode locking, machine learning)

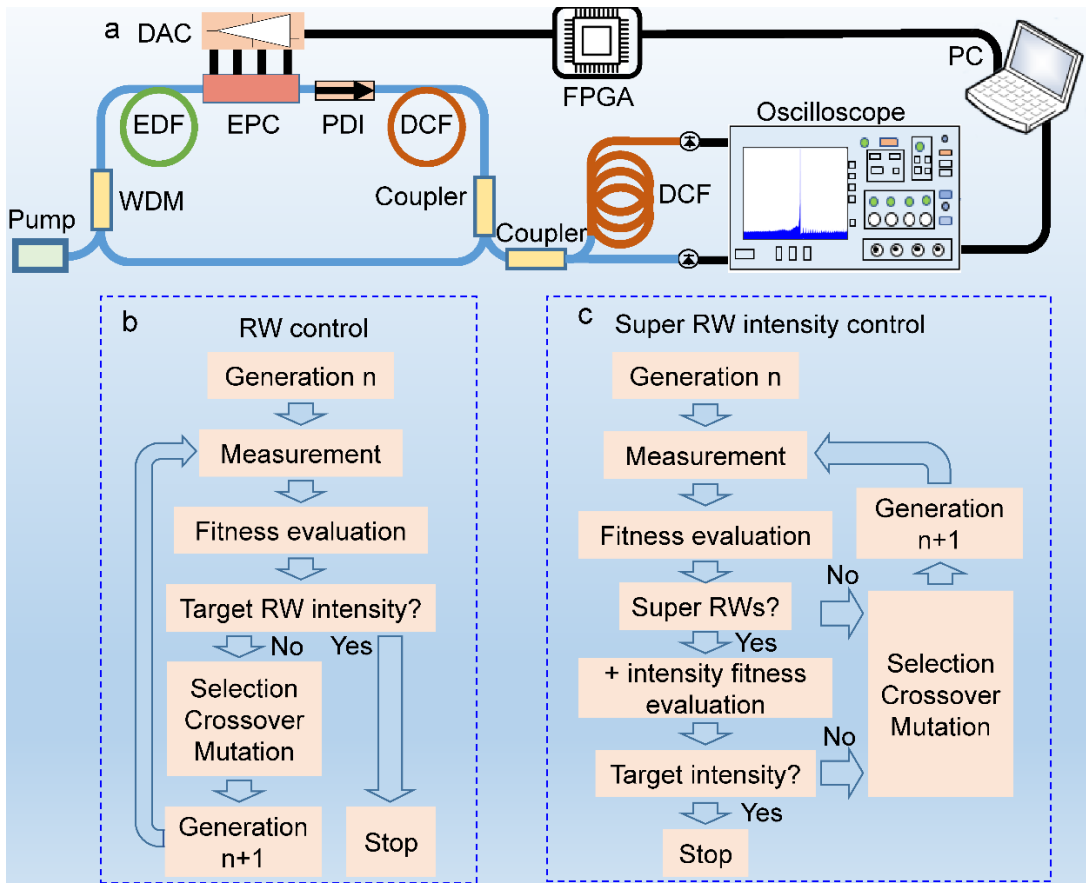

Fig. 1 (a) Schematic of the laser system. WDM: wavelength-division multiplexer; EDF: erbium-doped fibre; EPC: electronic polarisation controller; PDI: polarisation-dependent isolator; DCF: dispersion-compensating fibre; FPGA: field programmable gate array; DAC: digital-to-analogue converter. (b,c) Illustration of the EA principle for generating RWs with controllable intensity and controlling the intensity of super RWs, respectively.

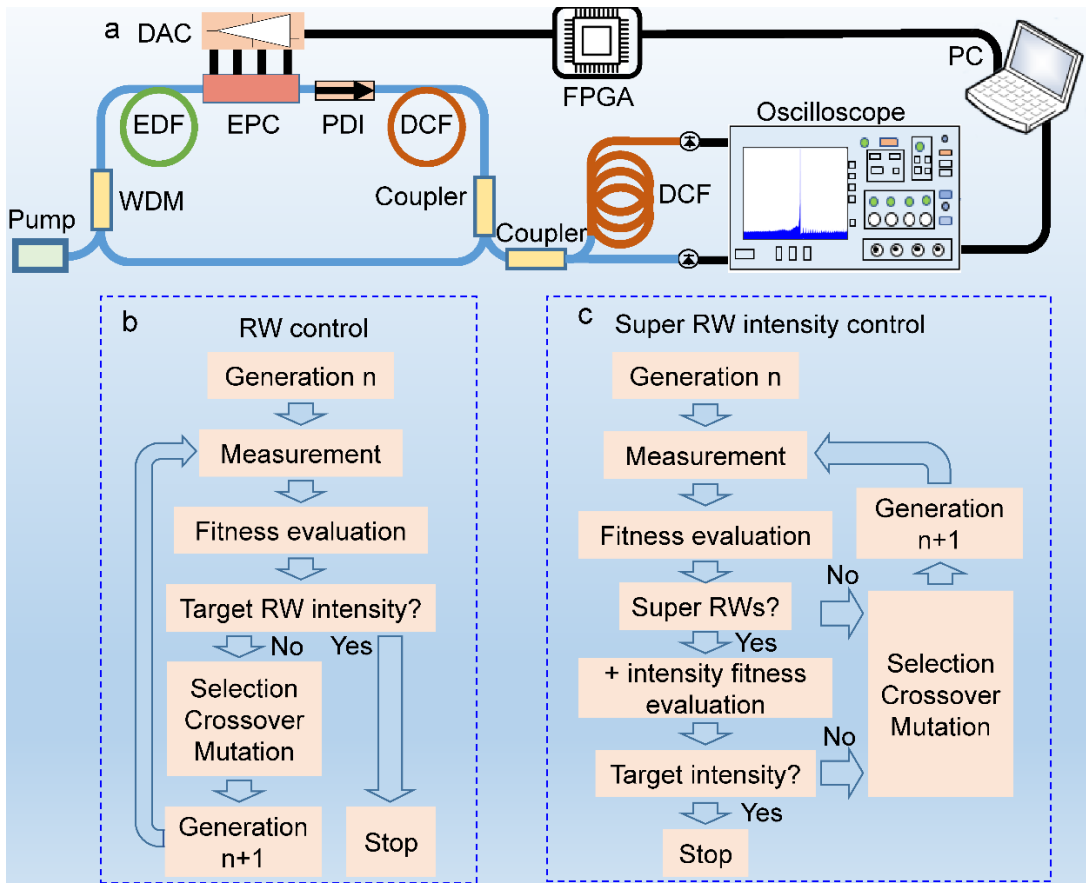

Fig. 1 (a) Schematic of the laser system. WDM: wavelength-division multiplexer; EDF: erbium-doped fibre; EPC: electronic polarisation controller; PDI: polarisation-dependent isolator; DCF: dispersion-compensating fibre; FPGA: field programmable gate array; DAC: digital-to-analogue converter. (b,c) Illustration of the EA principle for generating RWs with controllable intensity and controlling the intensity of super RWs, respectively.

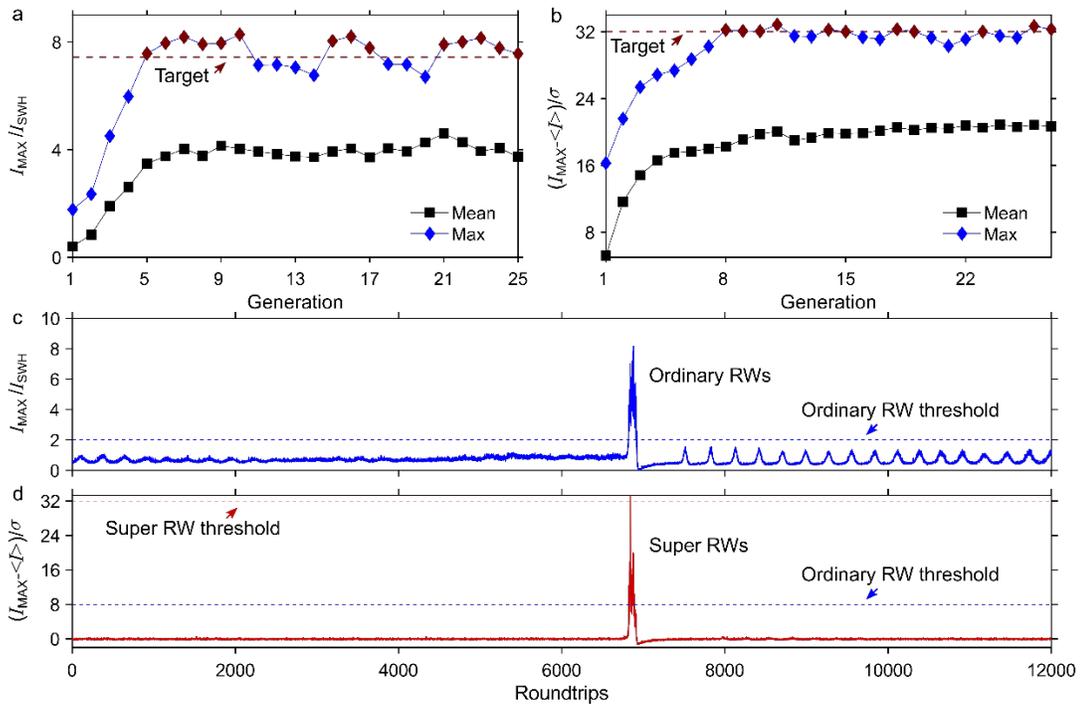

Fig. 2 (a, b) Evolutions of the average (squares) and maximum (diamonds) values of the ratios $I_{MAX}/I_{SWH}$ and $(I_{MAX}-<I>)/\sigma$ for ordinary and super RW optimisation, respectively, over successive generations. The dashed lines denote the respective target fitness ratios $C_{RW}$ and $C_{SRW}$. The blue diamonds (between the red diamonds) depict the generations that no longer meet the target fitness ratio; nevertheless, the target RW is later restored by the EA. (c, d) DFT recordings of the intensity maxima $I_{MAX}$ over 12,000 successive cavity round trips for the optimised ordinary and super RW laser operations. The blue and red dashed lines denote the peak intensity thresholds for ordinary and super RWs, respectively.

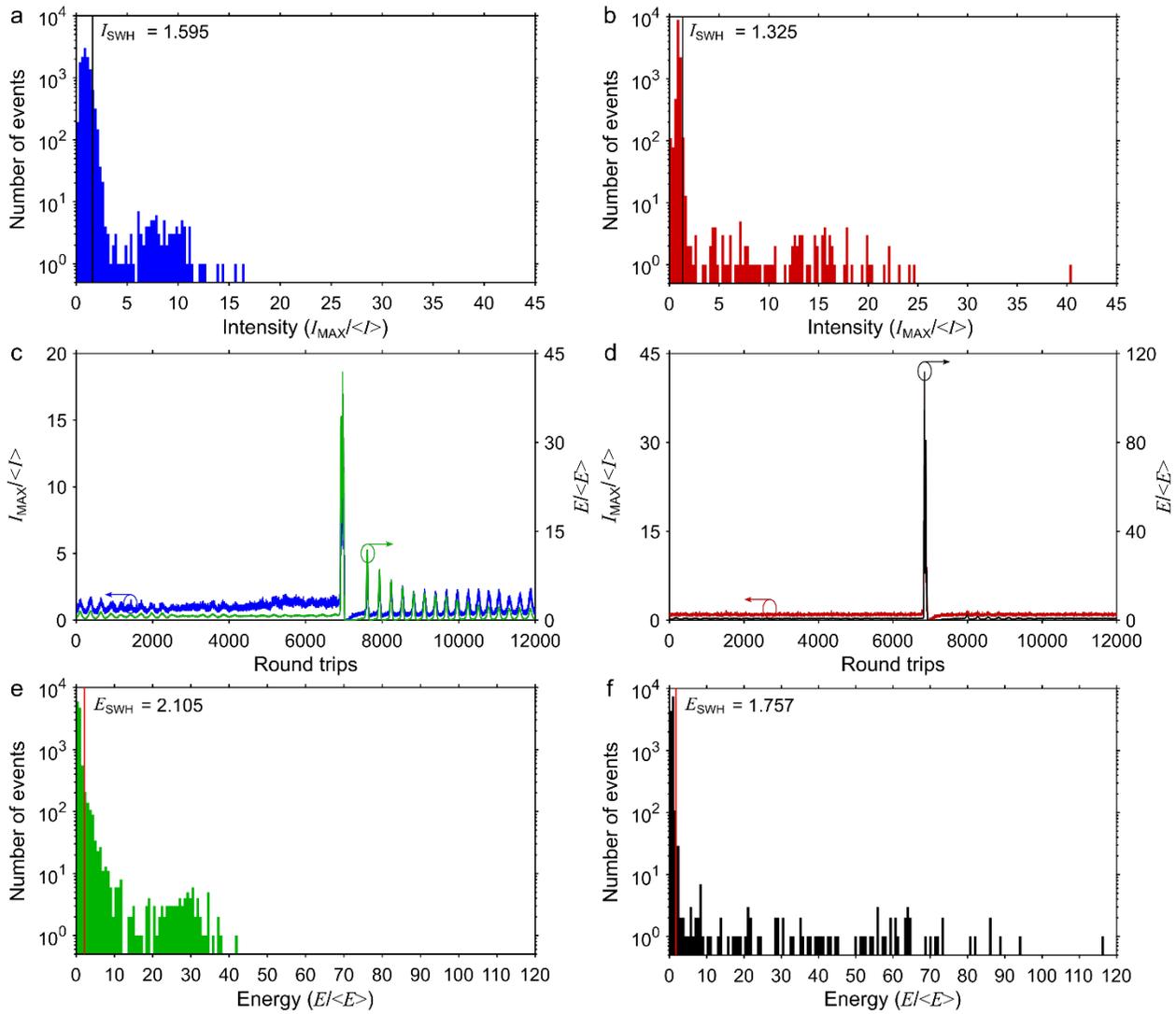

Fig. 3 (a,b) Histograms showing the distributions of the spectral intensity maxima for 12,000 successive cavity round trips, for ordinary and super RW laser operations, respectively. The black lines denote the associated significant wave heights ($I_{SWH}$). (c,d) Evolutions of the spectral intensity maxima (blue and red curves) and pulse energies (green and black curves) over 12,000 successive roundtrips for ordinary and super RW laser operations, respectively. (e,f) Corresponding histograms of the energy of the pulses for ordinary and super RW laser operations, respectively. The red lines denote the associated significant wave heights ($E_{SWH}$). In panels (c)-(f), $<E>$ represents the average energy within a roundtrip.

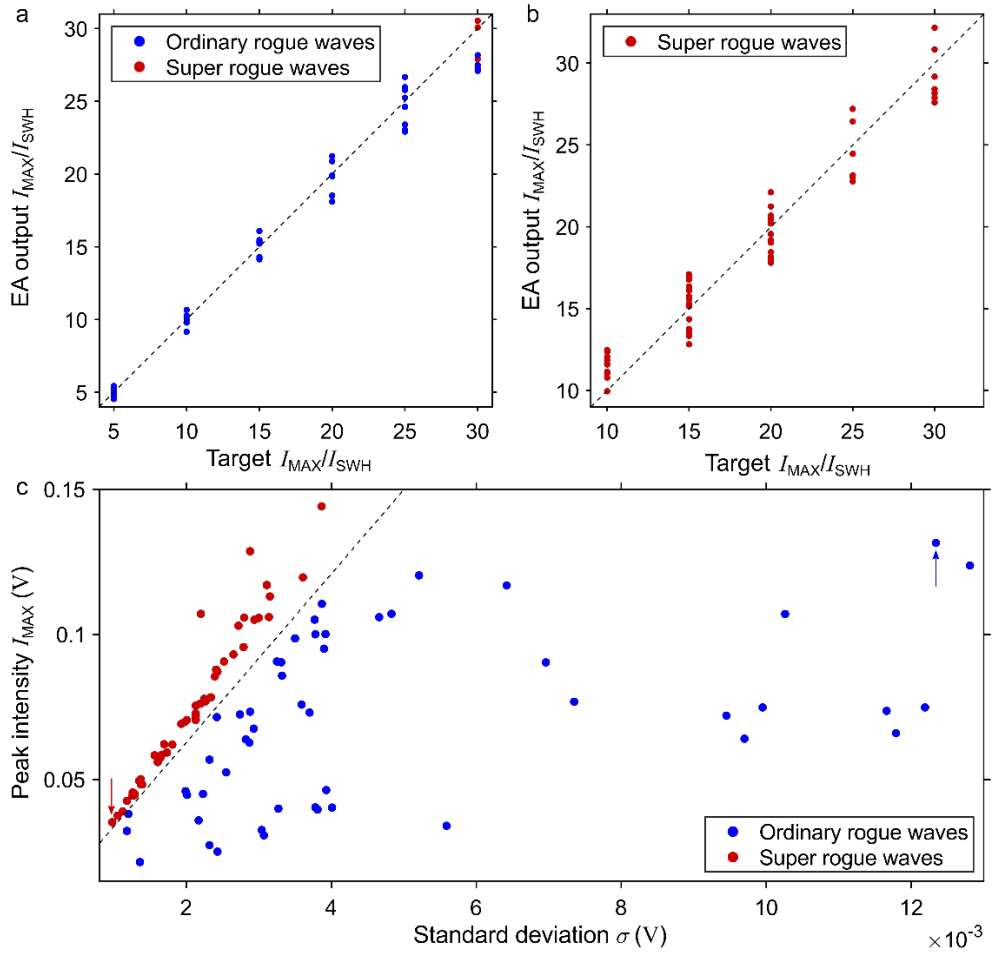

Fig. 4 (a) EA-based tuning of the strength of spectral RWs relying on the optimisation of the ratio $I_{MAX}/I_{SWH}$ (merit function $F_{RW}$): Regression between EA output and target $I_{MAX}/I_{SWH}$ values. Although the generated RWs can have large intensities, only three of them fulfil the definition of super RWs (red circles). (b) EA-based tuning of the strength of spectral super RWs relying on the joint optimisation of the ratios $I_{MAX}/I_{SWH}$ and $(I_{MAX} - \bar{I})/\sigma$ (composite merit function $F_{TSRW}$): Regression between EA output and target $I_{MAX}/I_{SWH}$ values. (c) Distribution of ordinary and super RWs in the plane of peak intensity $I_{MAX}$ and standard deviation $\sigma$. The blue and red arrows point at the strongest ordinary RW and the weakest super RW, respectively.

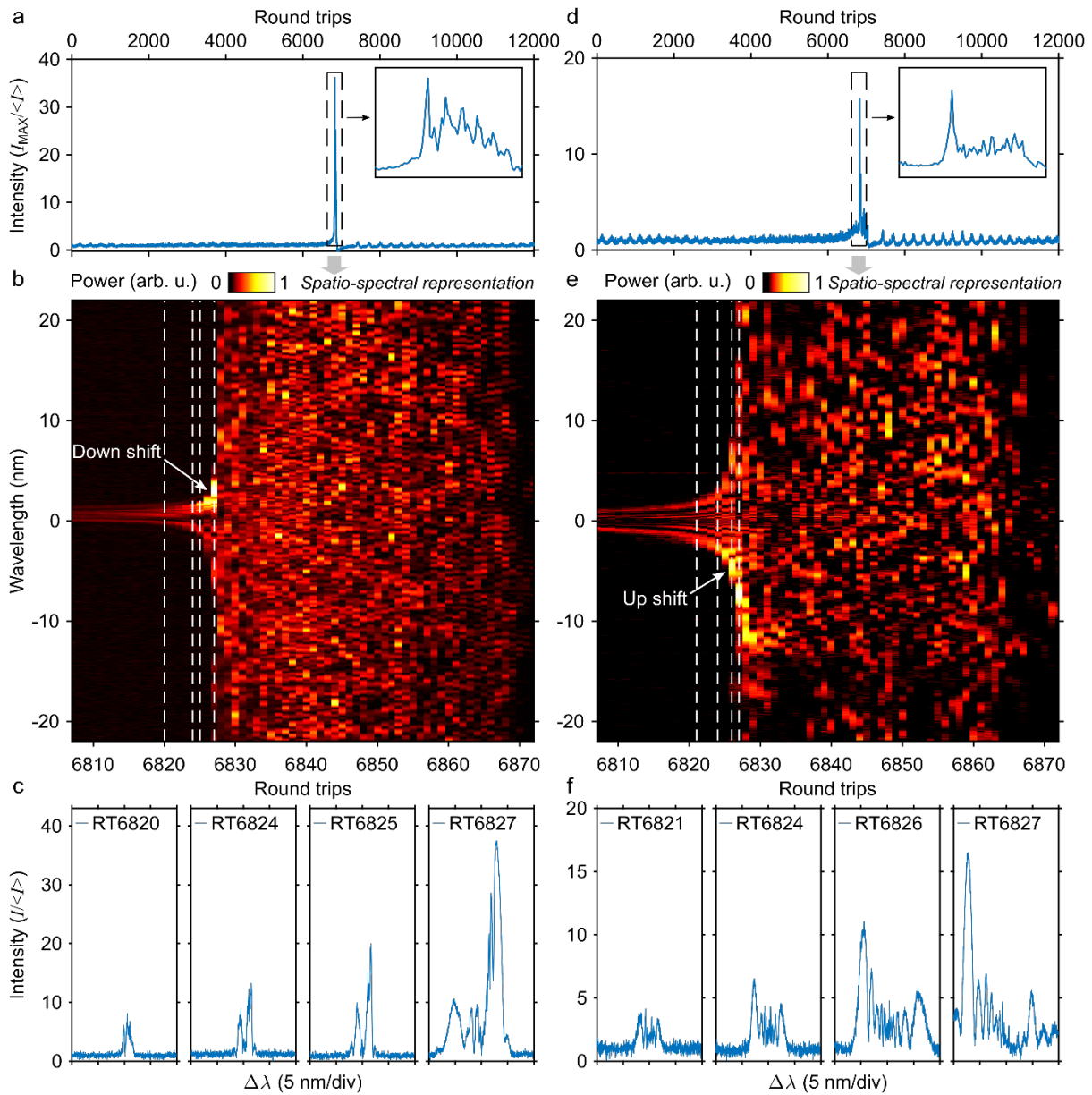

Fig. 5 Experimentally observed RW dynamics induced by frequency down-shifting and up-shifting. (a,d) Evolutions of the intensity maxima of the photo-detected signals after time stretching over $12\times10^3$ successive cavity round trips. The insets show magnified versions of the dashed rectangular regions. (b,e) Spatio-spectral representations of the laser intensity evolutions shown in the insets of panels (a,d), revealing the RW evolution dynamics. (c,f) Single-shot spectra at the round-trip numbers indicated by dashed lines in panels (b,e), highlighting the intensity growth of the down-shifted and up-shifted frequency components.

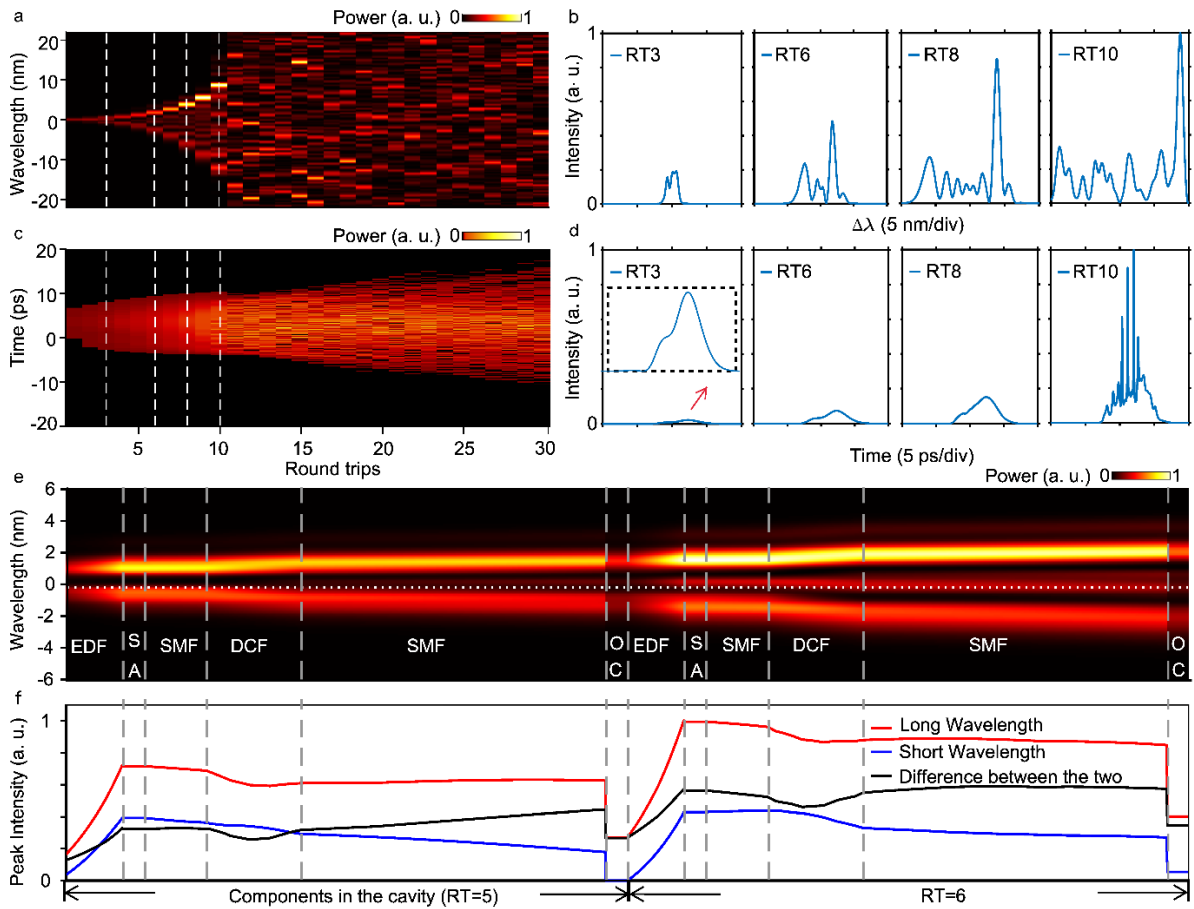

Fig. 6 Dynamics of frequency down-shifting induced RWs as obtained from numerical simulations of the laser model. (a) Evolution of the spectral intensity over successive cavity round trips. (b) Spectral intensity profiles at the round-trip numbers indicated by dashed lines in (a). (c) Corresponding round-trip evolution of the temporal intensity. (d) Temporal intensity profiles at the same round-trip numbers as in (b). (e) Intra-cavity evolution of the spectral intensity recorded over two consecutive traversals of the cavity (round-trip numbers 5 and 6), showing that the gain fibre is responsible for magnifying the energy difference between the long and short wavelength components of the spectrum. (f) Corresponding intra-cavity evolution of the intensity of the spectral peak. SA: saturable absorber, OC: output coupler.